\documentclass{JHEP}
\usepackage{amssymb}

\newcommand{\bea}{\begin{eqnarray}}
\newcommand{\eea}{\end{eqnarray}}
\newcommand{\bean}{\begin{eqnarray*}}
\newcommand{\eean}{\end{eqnarray*}}

\def\O #1{\overline{#1}}

\def\det{\mathop{\rm det}}

\def\d{{\rm d}}
\def\wt{\widetilde}

\def\th{{\theta}}

\def\a{{\alpha}}

\def\da{{\dot{\alpha}}}
\def\b{{\beta}}
\def\db{{\dot{\beta}}}

\preprint{
SNUST 030801\\
{\tt hep-th/0308049}}
\title{
Wilsonian Proof for Renormalizability \\
of ${\cal N}={1 \over 2}$ Supersymmetric Field Theories
\footnote{This work was supported in part by DOE grant
DE-FG02-90ER40542 (D.B.), by the KOSEF Interdisciplinary Research
Grant 98-07-02-07-01-5, and by the KOSEF Leading Scientist Grant
(S.-J.R).}}
\author{David Berenstein ${}^a$ , \,  Soo-Jong Rey ${}^{a,b}$\\
~~~~~~~~~~~~~~\\
${}^a$ Institute for Advanced Study\\
 Einstein Drive, Princeton NJ 08540 USA\\
 ~~~~~~~~~~~~~~~~\\
${}^b$ School of Physics \& BK-21 Physics Division\\
Seoul National University, Seoul 151-747 KOREA\\
~~~~~~~~~~~~~~~~~~~\\
 \email{dberens@ias.edu \qquad sjrey@gravity.snu.ac.kr} }
\abstract{We provide Wilsonian proof for renormalizability of
four-dimensional quantum field theories with ${\cal N}=1/2$
supersymmetry. We argue that the non-hermiticity inherent to these
theories permits assigning noncanonical scaling dimension both for
the Grassman coordinates and superfields. This reassignment can be
done in such a way that the non(anti)commutativity parameter is
dimensionless, and then the rest of the proof ammounts to power
counting. The renormalizability is also stable against adding
standard four-dimensional soft-breaking terms to the theory.
However, with the new scaling dimension assignments, some of these
terms are not just relevant deformations of the theory but become
marginal.}
\begin{document}
\section{Introduction}
Recently, deformations of superspace geometry have attracted
renewed attention, partly motivated by the quest to better
understand Ramond-Ramond background in string theory
\cite{deboer,OV, seiberg, berkovitsseiberg}. For instance, in type
IIB superstring compactified on a Calabi-Yau threefold $X$ one can
consider the dynamics of space-filling branes in the transverse
direction to the Calabi-Yau geometry in the presence of a
self-dual graviphoton flux on $\mathbb{R}^4$. Under these
conditions the ${\cal N}=1$ superspace is deformed to ${\cal N}={1
\over 2}$ superspace, whose chiral and antichiral Grassman-odd
coordinates obey the Clifford and the Grassman algebras,
respectively,
\bea \{ \th^\a, \th^\b \} = C^{\a\b}, \qquad \{\O \th^\da, \O
\th^\db \} = 0, \qquad [y^m, y^n] = 0, \label{nac} \eea
where $C^{\a\b}$ refers to the self-dual graviphoton flux measured
in units of string scale and $(y^m, \th^\a, \O \th^\da)$ denotes
the ${\cal N}=1$ superspace coordinates in the chiral basis.
Accordingly, once the graviphoton background is turned on,
The Euclidean worldvolume dynamics of the D3-branes gets modified, and
is governed by a non(anti)commutative gauge theory with ${\cal N}={1
\over 2}$ supersymmetry.

With such motivations, various aspects of field theories defined
on the deformed superspace have been studied extensively
\cite{seiberg, brittofengrey1, jungtay, brittofengrey2, grisaru,
brittofeng, italian, luninrey}. It was pointed out \cite{seiberg}
that the deformation (\ref{nac}) induces local operators
multiplied by the non(anti)commutativity parameter $C^{\a\b}$.
These induced operators are typically of higher scaling dimension,
and they might well render the deformed theories
nonrenormalizable. Surprisingly, it turned out the deformed
theories are renormalizable to all orders in perturbation theory.
The first important observation was that the deformed Wess-Zumino
model is renormalizable up to two loops \cite{grisaru}. The
renormalizability was then extended to all orders in perturbation
theory for the deformed Wess-Zumino model in \cite{brittofeng,
italian}, and for deformed gauge theories with(out) matter in
\cite{luninrey}. In particular, these works show that, though they
carry scaling dimensions larger than four, the deformation-induced
operators are radiatively corrected at most logarithmically.

In this work, we offer an intuitive proof for renormalizability of
deformed quantum field theories with ${\cal N}={1 \over 2}$
supersymmetry. The key observation is that the deformation
(\ref{nac}) is chirally asymmetric and renders operators induced
by the deformation non-hermitian. This implies that a field theory
defined on deformed superspace is non-unitary, so we might as well
relax or drop out other assumptions normally required for unitary
field theories from the very beginning. The main idea is then to
use a different scaling dimension for the superspace coordinates
and superfields that appear in the supersymmetric Lagrangian so
that, with the new scaling dimension assignment, perturbations
around a Gaussian fixed point are marginal or relevant by
power-counting. Furthermore, we show that there are only finitely
many relevant and marginal operators, so this constitutes a proof
of perturbative renormalizability. Any notion and meaning that
should be assigned to the new dimensional analysis at
nonperturbative level will not be addressed.

This work is organized as follows. In section 2, we study general
operator analysis in deformed non(anti)commutative field theories,
and provide Wilsonian proof for the renormalizability. In section
3, we discuss various points worthy of mention. we show that the
renormalizability is stable against adding soft-breaking operators
and variant choices of the superpotentials. We also draw analogy
and comparison to known nonunitary conformal field theories.
\section{Intuitive Proof of Renormalizability}
\subsection{Standard and non-standard dimensional analysis}
Consider a ${\cal N}=1$ supersymmetric quantum field theory,
described by the Lagrangian $L_0$, so it consists of D-, F-, and
$\O {\rm F}$-terms. After the deformation (\ref{nac}), the
deformed theory has ${\cal N}={1 \over 2}$ supersymmetry only, and
is described by the Lagrangian of the form:
\bea L = L_0({\cal O}, {\cal O}^\dagger) + \sum_{i} \wt {\cal
O}_i(C). \label{deformedL} \eea
The first part, $L_0$, is the Lagrangian of the undeformed theory
consisting of local operators ${\cal O}, {\cal O}^\dagger$. The
remainder consists of local operators $\wt {\cal O}_i$ induced by
the deformation (\ref{nac}). We include the explicit dependence on
the non(anti)commutativity parameter $C^{\a\b}$ into the
definition of these operators. Typically, the deformation-induced
operators are operators of dimension four or higher, and the
operators $\wt {\cal O}_i$ are suppressed in the (anti)commutative
limit $C^{\a\b} \rightarrow 0$. Other than this, for the foregoing
discussions, explicit form of the operators is not needed.

Quite surprisingly, though involving higher-dimensional operators,
such deformed theories exhibits perturbative renormalizability. It
was shown, by explicit computation or by operator analysis and
power-counting, that both Wess-Zumino model \cite{grisaru,
brittofeng, italian} and gauge theories \cite{luninrey} defined on
the deformed superspace are renormalizable. Certainly, this is not
a feature inherent to ordinary quantum field theory, and calls for
a better way of understand this issue.

An important clue is provided by the observation that the
deformation parameter $C^{\a\b}$ in (\ref{nac}) acts as the
R-symmetry-breaking parameter. If we start with the undeformed ${\cal
N}=1$ theory with a U(1) R-symmetry \footnote{Here, the R-symmetry
may not be an exact symmetry but is broken by various couplings.
In this case, by R-symmetry, we refer to (pseudo) R-symmetry in
which R-symmetry breaking parameters are promoted to appropriate
superfields.}. That is, under R-symmetry
\bea \th^\a \rightarrow e^{+ i \delta} \th^\a, \qquad \O \th^\da
\rightarrow e^{-i \delta} \O \th^\da, \nonumber \eea
the non(anti)commutativity parameter carries R-charge $-2$:
$C^{\a\b} \rightarrow e^{ - 2 i \delta} C^{\a\b}$. We thus
interpret $C^{\a\b}$ as a spurion of R-charge -2. We also recall
that, in the ordinary ${\cal N}=1$ superspace, from the analytic
 continuation from Lorentzian spacetime $\mathbb{R}^{3,1}$,
$\th^\a$ and $\O \th^\da$ carry equal scaling dimensions,
 so
\bea [ \th^\a ] = -{1 \over 2}, \qquad [\O \th^\da ] = -{1 \over
2}, \qquad [y^m ] = - 1. \eea
In Lorentzian spacetime $\mathbb{R}^{3,1}$, the first two follow
from the unitarity requirement of a given theory, viz. $\O \th^\da
= (\th^a)^\dagger$. In Euclidean spacetime $\mathbb{R}^4$, we
ordinarily continue working with the same scaling dimensions of
the Grassman-odd coordinates. With such R-symmetry and scaling
dimension assignments, the D-term in the Lagrangian is always
hermitian, while the $\O {\rm F}$-term is hermitian conjugate to
the F-term. Likewise, real superfield $V$ continues to be real,
and chiral superfield $\Phi$ and antichiral superfield $\O \Phi$
are hermitian conjugate each other. Moreover, local operators
${\cal O}$ and their hermitian-conjugate operators ${\cal
O}^\dagger$ carry the same scaling dimensions:
\bea [{\cal O}] = \Delta \quad \longleftrightarrow \quad [{\cal
O}^\dagger] = \Delta, \eea
and all of these results are true from unitarity considerations.

For the theories under consideration, we are in a different and
interesting situation. Since the non(anti)commutative deformation
treats $\th^\a$ differently and independently from $\O \th^\da$,
the deformed theories are defined only in Euclidean spacetime
$\mathbb{R}^4$. Wick rotation to Lorentzian spacetime
$\mathbb{R}^{3,1}$ is not permitted: first, it violates the Jacobi
identities of the non(anti)commutative deformations, and, second,
it is inconsistent with the self-duality of the graviphoton
background in the context of Type II string theory \footnote{The
associated graviphiton background is either complex or it carries
energy-momentum and there is back-reaction to the background
geometry.}. As chiral and antichiral coordinates are a priori
independent, various properties implicit to ordinary superspace no
longer need to hold. For example, a vector superfield $V$ becomes
complex-valued because of the $C^{\a\b}$-dependence:
\bea V(y, \th, \O \th) = V_{\rm ordinary} (y, \th, \O \th) - {i
\over 4} \O \th \O \th \th^\a {C_\a}^\gamma \sigma^m_{\gamma \dot
\gamma} \{\O \lambda^{\dot \gamma}, A_m\}. \nonumber \eea
A chiral superfield $\Phi(y, \th)$ and antichiral superfield $\O
\Phi(\O y, \O \th)$ are independent each other (except for the
fact that they are paired in the kinetic term), and so are the
superpotentials $W(\Phi)$ and $\O W(\O \Phi)$. Thus, in the
absence of hermiticity property, we can think of the fields $\Phi$
and $\O \Phi$ as chiral and antichiral superfields, independent
each other, and assign non-standard but perfectly sensible scaling
dimension to the chiral coordinates $\th^\a$ different from that
for antichiral coordinates $\O \th^\da$. Explicitly, with a
continuous parameter $\delta$, we can assign
\bea [\th^\a]  = -{1 \over 2} + \delta, \qquad [\O \th^\da] = -{1
\over 2} - \delta, \qquad [y^m] = -1, \label{newassignment}\eea
and, for the non(anti)commutativity parameter, $[C^{\a\b}]= - 1 +
2 \delta$. Notice that the new scaling dimension assignment is
compatible with the anticommutation relations \footnote{Notice
that because $\O Q_\da$ is not a conserved charge and the third
relation is not a derivation operation, these equations do not
form a Lie algebra.} involving ${\cal N}={1 \over 2}$ supercharges
$Q_\a$:
\bea \{ Q_\a, Q_\b \} &=& 0 \nonumber \\
\{ Q_\a, \O Q_\da \} &=& \sigma^m_{\a \da} P_m \nonumber \\
\{ \O Q_\da, \O Q_\db \} &=& \sigma^m_{\a \da} \sigma^n_{\b \db}
P_{m} P_n, \eea
where $[Q_\a] = 0, [\O Q_\da] = +1$ and $[P_m] = +1$. We can also
assign different scaling dimension to chiral superfields
differently from their corresponding antichiral superfields, and
scaling dimension of a chiral operator differently from that to
its hermitian-conjugate antichiral operator. Possibility of the
nonstandard scaling dimension assignment (\ref{newassignment})
constitutes the crux for proving renormalizability of ${\cal N}={1
\over 2}$ supersymmetric field theories defined on deformed
superspace (\ref{nac}).

\subsection{Renormalization group flow}
With our nonstandard scaling dimension assignment, we can explain
intuitively renormalizability of the non(anti)commutative field
theories. Recall that, in the Wilsonian approach \cite{wilson},
renormalizability is ensured if all operators involved in the
renormalization group flow retain scaling dimensions equal to four
or less {\sl and} if there are only a finite number of such
operators. More precisely, renormalizability follows if flows of
renormalized couplings form a finite-dimensional subspace in the
infrared in the total space of all possible operators.

Notice that the Wilsonian renormalization group flow, as
summarized above, does not rely explicitly on how one assigns
scaling dimension to each elementary field and each Grassman-odd
coordinate of the superspace. In particular, if we assume we have
a Gaussian fixed point in the ultraviolet, the perturbative
expansion will be given by the same Feynmann diagrams with any
other assignment of the scaling dimensions. It is thus
advantageous to assign scaling dimensions to be the most
convenient of all possible choices. Such a freedom is not
available for theories with hermiticity. In the present context,
the full theory is given by (\ref{deformedL}) and is
non-hermitian. Since the deformation-induced operators $\wt {\cal
O}_i$ carry dimensions higher than four in the standard scaling
dimension assignment, we would look for a new assignment wherein
these induced operators (along with the Gaussian terms) carry
scaling dimension four. Such assignment is motivated in part by
the observation \cite{grisaru,brittofeng,italian,luninrey} that
possible radiative corrections to these operators are logarithmic
only. It is not hard to see that the most suitable choice is
$\delta = 1/2$ in (\ref{newassignment}), viz.
\bea \delta = {1 \over 2} \quad : \quad\quad [\th^\a] = 0, \qquad
[\O \th^\da] = -1, \qquad [y^m] = -1. \label{new} \eea
It immediately follows that the non(anti)commutativity $C^{\a\b}$
parameter can be promoted to a {\sl dimensionless} coupling
parameter. This means that for the free field theory the
non(anti)commutative deformation does not introduce any scaling
violations, and can still be considered as an ultraviolet fixed
point. This should be sharply contrasted against the
noncommutative field theories, where the deformation parameter
$\Theta^{mn}$ carries always dimensions and spoils the idea that,
in the ultraviolet, there is a Wilsonian renormalization-group
fixed point, from which one is perturbing away to obtain the
infrared dynamics.

The new scaling dimension assignment leads to the following virtue
when a theory is deformed by the non(anti)commutativity
(\ref{nac}). As $[\th^\a]$ and $[C^{\a\b}]$ are assigned
dimensionless, the star product,
\bea \star = \exp \left( -{1 \over 2} C^{\a\b}
\overleftarrow{Q_\a} \overrightarrow{Q_\b} \right) \qquad {\rm
where} \qquad Q_\a := {\partial \over \partial \th^\a}, \nonumber
\eea
used for the non(anti)commutative deformation (\ref{nac}) would
not generate operators with higher scaling dimensions. Therefore,
once the undeformed theory is renormalizable (with the new
counting of dimension), the new theory deformed with
non(anti)commutative is manifestly renormalizable as well.

\subsection{Wess-Zumino model}
This much said, we now examine explicitly the lagrangian for the
deformed Wess-Zumino model and test our argument further.

Consider the deformed Wess-Zumino model with
\bea K_\star (\Phi, \O \Phi) = \O \Phi \Phi, \qquad W_\star (\Phi)
= ({m \over 2} \Phi^2 + {g \over 3} \Phi^3)_\star, \qquad \O
W_\star (\O \Phi) = ({\O m \over 2} \O \Phi^2 + {\O g \over 3} \O
\Phi^3)_\star. \eea
The undeformed theory is unitary and renormalizable. With the
deformation (\ref{nac}), the theory is a sum of ordinary
Wess-Zumino model with ${\cal N}=1$ supersymmetry and a
deformation-induced operator \cite{seiberg}
\bea \wt {\cal O} = -{g \over 3} |C| F^3, \eea
where $|C| \equiv \det C^{\a\b}$. Evidently, $\wt {\cal O}$ is
nonhermitian, and the theory is nonunitary.

Now, we can make the dimensional analysis with the new scaling
dimension assignments for $\theta^\a, \O \theta^\da$. The D-term
measure $\int \d^2\theta \d^2\bar\theta$ has the scaling dimension
$2$, viz. its dimension does not change. The $\mathbb{R}^4$
measure $\d^4 y$ has scaling dimension $-4$. The kinetic term
should be marginal in the ultraviolet, so $[\Phi]=2-[\O \Phi]$.
The F-term measure $\int \d^2\theta$ is dimensionless, while the
$\O {\rm F}$-term measure $\int \d^2\bar\theta$ has scaling
dimension $2$. In order for the cubic term in the undeformed
antiholomorphic superpotential $\O W (\O \Phi)= \O \Phi^3$ to be
at most marginal, we need to assign scaling dimensions at the
fixed point such that $[\O \Phi]\leq 2/3$. From this, $[\Phi]\geq
4/3$, so the undeformed holomorphic superpotential $W(\Phi)=
\Phi^3$ is marginal only if $[\Phi]\leq 4/3$. It follows that we
need to assign the scaling dimensions such that $[\Phi]=4/3$ and
$[\O \Phi]= 2/3$. It also follows that all of their component
fields have positive scaling dimensions.

With these dimensions all of the possible terms in the lagrangian
have dimension less than or equal to four. Moreover, the mass
parameters $m, \O m$ have scaling dimensions $4/3, 2/3$,
respectively, so one still can promote them as the lowest
components of (anti)chiral superfields. Since both $\Phi, \O \Phi$
and all of their superfield components have positive dimensions,
any local operator involving polynomial of them and superspace
derivatives carries always a positive scaling dimension, and hence
there are only finitely many such operators with scaling dimension
less than or equal to four.

\subsection{Supersymmetric gauge theories}
Consider next the deformed gauge theories. The undeformed part of
the theory is the ordinary ${\cal N}=1$ supersymmetric gauge
theory with massless matter. The Lagrangian has no dimensionful
coupling parameter and has U(1) R-symmetry. The
deformation-induced operators consist of \cite{seiberg, ito}
\bea \wt{\cal O}_1 &=& C^{\a\b} {\rm Tr} \{ F_{(\a\b)} \O \lambda
\O \lambda\}, \qquad \quad \wt {\cal O}_2 = |C|^2 {\rm Tr} (\O
\lambda
\O \lambda)^2 \nonumber \\
\wt {\cal O}_3 &=& C^{\a\b} (\nabla_m \O \varphi) \sigma^m_{\a\da}
\O \lambda^\da \psi_\b, \qquad \wt {\cal O}_4 = C^{mn} \O \varphi
F_{mn} F, \qquad \wt {\cal O}_5 = |C|^2 \O \varphi \O \lambda \O
\lambda F. \label{gaugethops} \eea
Again, with $[\th^\a]=0$ and $[\O \th^\da]=-1$, the superfield
analysis yields that $[W_\a]=2$ and $[\O W_\da]= 1$. It assigns
scaling dimension unchanged for the vector field, $[V_m]=1$, but
changed for the gauginos as $[\lambda]=2$, $[\bar\lambda]=1$. The
deformation of field strength superfields \cite{seiberg}
\bea W_\a &=& W_\a \Big|_{\rm ordinary} + C_{\a\b} \th^b \O
\lambda \O \lambda \nonumber \\
\O W_\da &=& \O W_\da \Big|_{\rm ordinary} - \O \th \O \th \left[
C^{\a\b} \{ F_{(\a\b)}, \O \lambda_\da \} + C^{mn} \{ A_m,
\nabla_n \O \lambda_\da - {i \over 4} [A_n, \O \lambda_\da]\} + {i
\over 16} |C|^2 \{ \O \lambda \O \lambda, \O \lambda_\da \}
\right] \nonumber \eea
is automatically compatible with the new scaling dimension
assignment and $[C^{\a\b}]= 0$. The new scaling dimensions for
matter superfields are the same as the Wess-Zumino model case.

With the nonstandard scaling dimension assignment, the undeformed
part of the Lagrangian contains again operators of dimension equal
to four only. The gauge coupling parameter remains dimensionless.
The deformation-induced operators are (\ref{gaugethops}), and they
all have new scaling dimensions equal to four. Hence, the entire
deformed Lagrangian contains operators of dimension four only and
the theory is manifestly renormalizable.

\section{Further Discussions}
\subsection{Soft-breaking terms}
We can also utilize the idea developed in the previous section,
and show that the standard soft-breaking terms in four-dimensional
supersymmetric field theories are relevant or at most marginal.
This is a simple exercise in dimensional analysis with the
nonstandard scaling dimension assignment.

The soft-breaking operators consist of the followings. Mass terms
for scalar operators are of scaling dimension $4/3$, $2$, and
$8/3$, respectively, while cubic terms can be of up to dimension
$4$ when we take the field of scaling dimension $4/3$ to the cubic
power. Also a gaugino mass term yields terms of dimension $4$ and
$2$ for chiral and antichiral components, respectively.

A surprising fact is that, with the new scaling dimension
assignment, some of the standard soft-breaking terms turn into
marginal ones. This is because these terms also induce breaking of
the R-symmetry, and then they are not protected by the
nonrenormalization theorems. Presumably, this would spoil various
relations among the renormalized couplings in the full theory and
one should instead consider the most general Lagrangian with
fields given as above, with all possible operators of dimension
less than or equal to four that can be made out of polynomials of
the fields and their superspace derivatives.

Also, since rotational invariance of $\mathbb{R}^4$ is broken from
SO(4) to a subgroup which contains only an ${\rm SU}_R$(2)$\times$
U(1) isometry subgroup, we should include various operators which
are not Lorentz invariant, but they should still respect the above
isometry group. These are the terms which can be generated by
turning on the non(anti)commutative deformations.

\subsection{Variants}
Extending previous discussion, we can assign $[\Phi]=1+\epsilon$
and $[\O \Phi]=1-\epsilon$, and so long as both dimension are
positive, there are only a finite number of relevant or marginal
operators in the field theory.

Recall that, for Euclidean supersymmetries, the superpotentials
$W(\Phi)$ and $\O W(\O \Phi)$ are independent input data of a
given theory. Denote the highest monomial of $W(\Phi)$ and $\O W
(\O \Phi)$ as $n, \O n$, respectively. In case $n = \O n$, by
repeating the dimensional analysis with the new scaling dimension
assignment, we observe that the theory is renormalizable only if
$n = \O n =3$ and $\epsilon=1/3$. This reproduces the standard
result. If $\O n = 2$, viz. $\O W(\O \Phi) = {\O m \over 2}
\O\Phi^2$, we can take $n$ up to four and the theory will still be
renormalizable. If $n=2$, viz. $W(\Phi) = {m \over 2} \Phi^2$,
then, surprisingly, we find that the dimension of $[\O \Phi]$ can
be made arbitrarily small and positive, so that the
antiholomorphic superpotential $\O W[\O \Phi]$ of any polynomial
is permitted while retaining renormalizability. In fact, because
of the ${\cal N}={1 \over 2}$ (non)renormalization theorem
\cite{brittofengrey1}, $\O W[\O \Phi]$ would not be renormalized
at all. Evidently, $\O W[\O \Phi]$ of arbitrary polynomial leads
to a richer structure of the ${\cal N}={1 \over 2}$ antichiral
ring.

Again, in all these variant cases, the non(anti)commutative
deformation would not spoil the renormalizability, as the
parameter $[C^{\a\b}]$ carries the scaling dimension zero.

\subsection{Analogy \& Comparison}
We should think of the pair of (anti)chiral superfields $\Phi, \O
\Phi$ or $W_\a, \O W_\da$  as the four-dimensional equivalent of
the $b,c$-system in two dimensions. Recall that the $b,c$-system
is ordinarily a non-unitary free field theory with the Lagrangian
$\int \d z \d \O z \, b \O \partial c$, and with $b$ and $c$
having different scaling dimension. The difference is not felt in
the correlators of the free fields themselves nor in the Feynman
rules if one works in the perturbation theory. However, the
energy-momentum tensor of the two-dimensional conformal field
theory carries an improvement term: $T = b\partial c - \alpha c
\partial b= T_{\rm ordinary} + a\partial J$, where $J$ is the U(1) ghost
current. Accordingly, scaling dimensions of various operators will
be modified depending on the assignment of the U(1) ghost charge
(the value of $a$). Drawing further analogy and comparison of the
$b,c$-system with the four-dimensional theories under discussion
might uncover surprises.

In this regard, the idea of nonstandard scaling dimension
assignment seems to match with the notion of twisting the
energy-momentum tensor by R-symmetry current that one does in
topological field theories \cite{tqft}. This may be an indication
that there exits a nonperturbative sector in non(anti)commutative
field theories which is topological in nature and for which the
deformed Lagrangian can be used to compute a certain class of
amplitudes or correlators. The antichiral ring (viz. operators
annihilated by $Q_\a$ of ${\cal N}={1 \over 2}$ supersymmetry) is
obviously part of it, and it would be interesting to understand if
the sector could be bigger than this.

\section*{Acknowledgement} We thank Ruth Britto, Sergey Cherkis,
Bo Feng, Akikazu
Hashimoto, Oleg Lunin, Herbert Neuberger, Nathan Seiberg and
Edward Witten for useful discussions. SJR was a Member at the
Institute for Advanced Study during this work. He thanks the
School of Natural Sciences for hospitality and for the
grant-in-aid from the Fund for Natural Sciences.

\end{document}